\documentclass[journal=nalefd,manuscript=letter,layout=onecolumn]{achemso}
\setkeys{acs}{email = false}

\usepackage{lineno}
\usepackage{hyperref}

\usepackage{chemformula} 
\usepackage[T1]{fontenc} 
\usepackage{graphicx}
\usepackage{amsmath}
\usepackage{siunitx}

\usepackage{titlesec}

\usepackage{graphicx}
\usepackage{xcolor} 
\usepackage{wasysym}
\usepackage{url}
\usepackage{amsmath}

\newcommand{\Figref}[1]{Fig.~\ref{#1}}
\newcommand{\Figrefa}[1]{Fig.~\ref{#1}(a)}
\newcommand{\Figrefb}[1]{Fig.~\ref{#1}(b)}
\newcommand{\Figrefc}[1]{Fig.~\ref{#1}(c)}

\newcommand{\Figrefe}[1]{Fig.~\ref{#1}(e)}

\author{Andrey A. Generalov}
\affiliation[VTT]
{VTT Technical Research Centre of Finland Ltd., P.O. Box 1000, FI‐02044 VTT, Espoo, Finland}
\altaffiliation{Authors to whom correspondence should be addressed: \href{mailto:andrey.generalov@vtt.fi}{andrey.generalov@vtt.fi},
\href{mailto:klaara.viisanen@vtt.fi}{klaara.viisanen@vtt.fi},
and \href{mailto:heorhii.bohuslavskyi@vtt.fi}{heorhii.bohuslavskyi@vtt.fi}}

\author{Klaara L. Viisanen}
\affiliation[VTT]
{VTT Technical Research Centre of Finland Ltd., P.O. Box 1000, FI‐02044 VTT, Espoo, Finland}
\altaffiliation{Authors to whom correspondence should be addressed: \href{mailto:andrey.generalov@vtt.fi}{andrey.generalov@vtt.fi},
\href{mailto:klaara.viisanen@vtt.fi}{klaara.viisanen@vtt.fi},
and \href{mailto:heorhii.bohuslavskyi@vtt.fi}{heorhii.bohuslavskyi@vtt.fi}}

\author{Jorden Senior}
\affiliation[VTT]
{VTT Technical Research Centre of Finland Ltd., P.O. Box 1000, FI‐02044 VTT, Espoo, Finland}

\author{Bernardo R. Ferreira}
\affiliation[VTT]
{VTT Technical Research Centre of Finland Ltd., P.O. Box 1000, FI‐02044 VTT, Espoo, Finland}

\author{Jian Ma}
\affiliation[Aalto University QCD, QTF]
{QCD Labs, QTF Centre of Excellence, Department of Applied Physics, Aalto University, P.O. Box 13500, FIN-00076 Aalto, Finland}

\author{Mikko Möttönen}
\affiliation[Aalto University QCD, QTF]
{QCD Labs, QTF Centre of Excellence, Department of Applied Physics, Aalto University, P.O. Box 13500, FIN-00076 Aalto, Finland}
\alsoaffiliation[VTT]
{VTT Technical Research Centre of Finland Ltd., P.O. Box 1000, FI‐02044 VTT, Espoo, Finland}

\author{Mika Prunnila}
\affiliation[VTT]
{VTT Technical Research Centre of Finland Ltd., P.O. Box 1000, FI‐02044 VTT, Espoo, Finland}

\author{Heorhii Bohuslavskyi}
\affiliation[VTT]
{VTT Technical Research Centre of Finland Ltd., P.O. Box 1000, FI‐02044 VTT, Espoo, Finland}
\altaffiliation{Authors to whom correspondence should be addressed: \href{mailto:andrey.generalov@vtt.fi}{andrey.generalov@vtt.fi},
\href{mailto:klaara.viisanen@vtt.fi}{klaara.viisanen@vtt.fi},
and \href{mailto:heorhii.bohuslavskyi@vtt.fi}{heorhii.bohuslavskyi@vtt.fi}}

\title{
Wafer-scale CMOS-compatible graphene Josephson field-effect transistors}

\begin{document}

\section*{Abstract}

Electrostatically tunable Josephson field-effect transistors (JoFETs) are one of the most desired building blocks of quantum electronics. JoFET applications range from parametric amplifiers and superconducting qubits to a variety of integrated superconducting circuits.
Here, we report on graphene JoFET devices fabricated with wafer-scale complementary metal-oxide-semiconductor (CMOS) compatible processing based on chemical vapour deposited graphene encapsulated with atomic-layer-deposited Al$_{2}$O$_{3}$ gate oxide, lithography defined top gate, and evaporated superconducting Ti/Al source, drain, and gate contacts.
By optimizing the contact resistance down to $\sim$ 170~$\Omega$\si{\micro\metre}, we observe proximity-induced superconductivity in the JoFET channels with different gate lengths of 150--350~nm.
The Josephson junction devices show reproducible critical current $I_{\text{C}}$ tunablity with the local top gate.
Our JoFETs are in short diffusive limit with the $I_{\text{C}}$ reaching  up to $\sim\,$3~\si{\micro\ampere} for a 50~\si{\micro\metre} channel width.
Overall, our demonstration of CMOS-compatible 2D-material-based JoFET fabrication process is an important step toward graphene-based integrated quantum circuits.
\\

The progress in studying mesoscopic physics and topological superconductivity, as well as applications of quantum information processing with proximitized semiconductor-superconductor hybrid devices, has thus far been mostly limited to chip-scale fabrication. 
The graphene devices have been identified as a promising candidate for realizing electric-field-tunable Josephson junctions (JJs) \cite{Li2016,Lee2018},
including exfoliated boron-nitride-encapsulated graphene devices \cite{Heersche2007,BenShalom2016,Borzenets2016,Butseraen2022}. Other investigated platforms for electric-field-tunable JJs are based on III-V and group-IV two-dimensional electron gas heterostructures, \cite{Barati2021,Aggarwal2021,Vigneau2019,Wen2021} and nanowires \cite{Xiang2006,Calado2015,Casparis2018,Phan2023}.
A particular attention has been focused on the development of Josephson field-effect transistors (JoFETs) where the proximity-effect-based supercurrent is controlled by an electrostatic gate electrode.
JoFETs enable a wide range of applications such as quantum coherent electronics, quantum memories, classical digital superconducting electronics, ultra-sensitive bolometers, non-reciprocal components, and quantum-limited parametric amplifiers  \cite{Casparis2018, Phan2023, Butseraen2022, Wen2019, Wen2021, Kokkoniemi2020,Sardashti2020}.

Recently, several important advancements in Chemical Vapour Deposition (CVD)-graphene-based scalable JoFET fabrication have been achieved, such as 150-mm-wafer CVD graphene JoFETs with a global back-gate \cite{Li2019} and the development of scalable van der Waals stacking of CVD graphene and hBN \cite{Schmidt2023}. 
However, the wafer-scale demonstration of CVD graphene JJs encapsulated with gate dielectric and with lithography defined top-gate control remains an important milestone to be demonstrated, essential for scaling the technology outside of the laboratory.

In this letter, we report scalable, lithography defined, electrically controlled superconducting proximity coupling for Al-graphene-Al JJs, for which the critical current $I_{\text{C}}$ (and thus kinetic inductance or Josephson energy) can be tuned by an order of magnitude with the local top-gate.

The JJs are fabricated using wafer-scale CMOS-compatible processes using CVD graphene wet-transferred on 150 mm Si wafer covered with 90 nm of SiO$_2$ thermal oxide (step $\#$1). For all process steps, patterning is performed using electron-beam lithography, and metal layers are deposited using the electron-beam evaporation.
First, the graphene is globally encapsulated with a layer of Al$_2$O$_3$ ($\#$2). Next, the source/drain (S/D) contacts are patterned, the Al$_2$O$_3$ at the contact area is wet-etched ($\#$3), and the S/D contacts are evaporated with a (5 nm/ 30 nm) Ti/Al stack ($\#$4). 
Then, a 30 nm of Atomic Layer Deposition (ALD) grown Al$_2$O$_3$ is deposited for gate dielectric ($\#$6).
Finally, the gate is patterned and the Ti/Al gate stack is evaporated ($\#$7). 
For the un-gated devices, also studied in this letter, only steps $\#$1-4 were carried out. 
The wafer-scale process used to fabricate JoFETs was optimized to reach the Al-graphene contact resistance down to $\sim\,$170 $\Omega\si{\um}$ (see \Figref {fig:fig2}, Table \ref{tbl:summary} and ref. \citenum{supp}), which is on par with the state of the art for top and edge-type graphene-metal contacts \cite{Li2016,Bonmann2019,Li2019,Xia2011,Ke2016}.
More details on fabrication are given in ref. \citenum{supp}. 

First, all fabricated graphene devices are probed at 300 K. To estimate the yield, we exclude the devices with shorted gates, or no gate tunability, or the total resistance normalized to $L_\text{g}$ > 1000~$\Omega$\si{\micro\metre} (indicating likely open circuit between Source and Drain contacts). Following these criteria, we obtain the device yield of $\sim$ 90\%. Then, the wafer is diced, and selected samples are cooled down in a dilution refrigerator with a base temperature of $T=42$~mK. 
DC and low-frequency lock-in measurements (AC current excitation is 35 nA$_{\text{rms}}$ at 33.3 Hz frequency) are carried out for the devices wire-bonded in two or four-probe configurations. In the former case, the wire resistance is subtracted (see the discussion in ref. \citenum{supp}).

\begin{figure}[t!]
\centering
\includegraphics[width=0.6\textwidth]{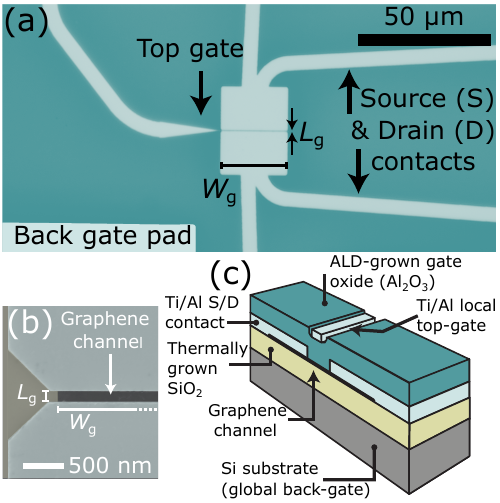}
\caption{
\textbf{(a)} False-color optical microscope image of a JoFET in a four-probe configuration
\textbf{(b)} False-color SEM image of the channel with length $L_{\text{g}}$ = 150~nm between the source and drain electrodes, taken before gate deposition.
\textbf{(c)} Three-dimensional schematic of the JoFET material stack.
}
\label{fig:fig1}
\end{figure}

\begin{figure}[t!]
\centering
\includegraphics[width=0.6\textwidth]{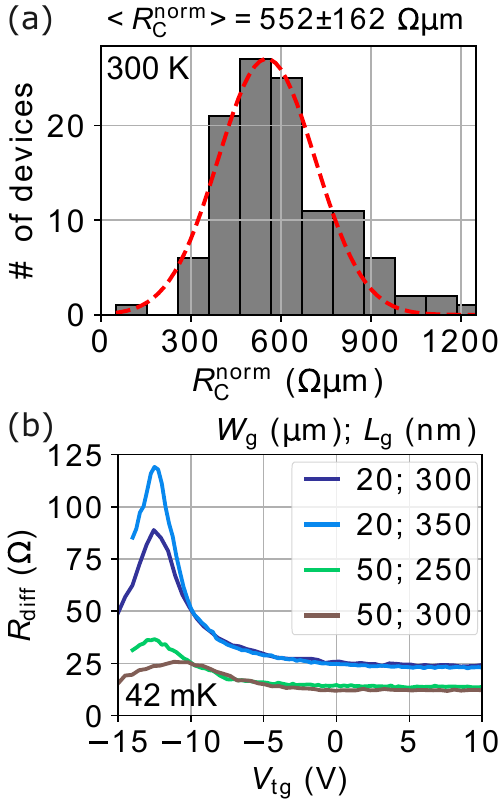}
\caption{
\textbf{(a)} Large-throughput room temperature characterization of contact resistance. 
\textbf{(b)} Normal-state differential resistance $R_{\text{diff}}$ as a function of top-gate voltage $V_{\text{tg}}$ for four JoFETs ($V_{\text{bg}} = 0$). Extracted Dirac points and contact resistances are given in Table \ref{tbl:summary}. 
}
\label{fig:fig2}
\end{figure}

\begin{figure*}[t!]
\centering
\includegraphics[keepaspectratio, width=0.9\textwidth]{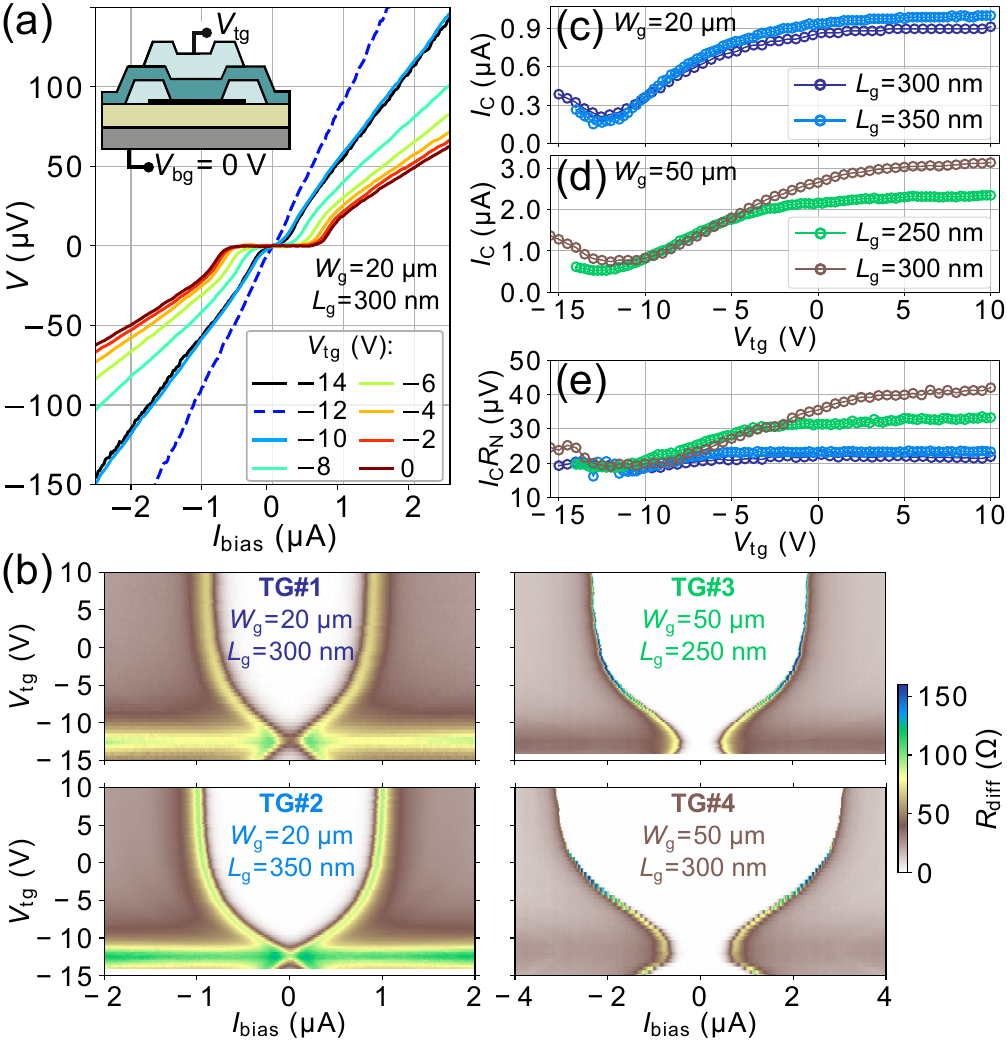}
\caption{
\textbf{(a)} Measured voltage across the junction $V$ as a function of direct-current bias $I_{\text{bias}}$ for $T = 42$~mK and the indicated top-gate voltages $V_{\text{tg}}$ (device TG$\#1$). Inset shows a simplified device sketch and the gate-biasing configuration. 
\textbf{(b} Measured differential resistance $R_{\text{diff}}$ as a function of $V_{\text{tg}}$ and $I_{\text{bias}}$ for two JoFET devices  with $W_{\text{g}}$ = 20~\si{\micro\m} (TG$\#1,2$) and two - with 50~\si{\micro\m} (TG$\#3,4$).
\textbf{(c,d)} Critical current $I_{\text{C}}$ extracted from \textbf{(b)} as a function of $V_{\text{tg}}$ for TG$\#1,2$ and TG$\#3,4$, respectively.
\textbf{(e)} The product of the critical current and normal-state resistance $I_{\text{C}}R_{\text{N}}$ calculated for TG$\#1-4$ using data in \textbf{(b)}. Same color code is used as in \textbf{(c,d)}. 
}
\label{fig:fig3}
\end{figure*}

The optical and scanning-electron-microscope (SEM) images of the final top-gated JoFET are shown in \Figref{fig:fig1}~(a,b), where the JoFET device channel width ($W_{\text{g}}$) and gate length ($L_{\text{g}}$) are indicated.
A schematic cross-section of the JoFET is given in \Figrefc{fig:fig1}.
The wafer-scale contact resistance characterization, being one of the key parameters for graphene JoFET device optimization, is shown in \Figrefa{fig:fig2}.
Based on the measurements of more than 100 devices with fixed $W_{\text{g}}$ = 20 \si{\micro\metre}  and $L_{\text{g}}$ from 8 \si{\micro\metre} to 150 nm and using the Gaussian fit of the data, we obtain the average contact resistance normalized to $W_{\text{g}}$ of $R^{\text{norm}}_{\text{C}}$ = 552 $\pm$ 162 $\Omega$\si{\micro\metre}.
\Figrefb{fig:fig2} shows the normal-state differential resistance vs top-gate voltage characteristics $R_{\text{diff}}(V_{\text{tg}})$ with an on$/$off ratio of $\sim\,$3--5  for four JoFETs with gate length $L_{\text{g}} = 250 - 350 $~nm and width of $W_{\text{g}} = 20$~\si{\micro\metre} and 50~\si{\micro\metre}. These measurements were performed at 42 mK using current biasing with $I_{\text{bias}}$ larger than JoFET critical current $I_{\text{C}}$ (see \Figrefb {fig:fig3}).

All measurements of the top-gated JJs are done by sweeping $V_{\text{tg}}$ from negative to positive voltages to avoid gate-hysteresis effects (see ref. \citenum{supp}).
We do not observe any significant current hysteresis as shown in ref. \citenum{supp}.
The devices are stable against thermal cycling in terms of $I_{\text{C}}$ and $V_{\text{Dir}}$ between different cooldowns (see ref. \citenum{supp}).

Following the work by Li et al.~\cite{Li2016} on graphene JJs and supported by the results in refs. \cite{Xia2011,Cayssol2009,CastroNeto2009}, we extract the contact resistance $R_{\text{C}}$, Dirac peak position $V_{\text{Dir}}$, and charge carrier concentration $n_{\text{c}}$ (using the model from ref. \citenum{Kim2009}) from the 42 mK data presented in \Figrefb{fig:fig2}. 
As summarized in Table \ref{tbl:summary}, the resulting $V_{\text{Dir}}\approx -11...-12$~V, 
normalized contact resistance $R^{\text{norm}}_{\text{C}} \approx 165 - 300$~$\Omega$\si{\micro\metre} for TG$\#1-4$.  
At $V_{\text{tg}} = 10$~V, we obtain the mean free path $l_{\text{e}} = 35-70$~nm from $L_{\text{g}}/(R_{\text{ch}}W_{\text{g}}) = 2e^{2}k_{\text{F}}l_{\text{e}}/h$, where $R_{\text{ch}}$ is the channel resistance (the total normal-state resistance is $R_{\text{ch}} + 2 R_{\text{C}}$), $e$ is the elementary charge, $k_{\text{F}}$ is the Fermi wave vector calculated as $\sqrt{\pi n_{\text{c}}}$, and $h$ is the Plank constant. 

The superconducting coherence length $\xi_{\text{s}} = 360 - 510$ nm (larger than $L_{\text{g}}$) is obtained from $\sqrt{\hbar D / \Delta}$, where $\hbar$ is the reduced Plank constant, $\Delta$ = 90 \si{\micro\eV} is induced superconducting gap extracted from multiple Andreev reflection data (see ref. \citenum{supp}), and $D$ is the diffusion coefficient calculated as $v_{\text{F}}l_{\text{e}}/2$, with $v_{\text{F}} \approx$ $10^{6}$ m/s being the Fermi velocity in graphene far from $V_{\text{Dir}}$ \cite{CastroNeto2009}.
Next, we deduce the Thouless energy $E_{\text{Th}} = 125 - 200$~\si{\micro\eV} (larger than $\Delta$) for the diffusive regime ($l_{\text{e}} < L_{\text{g}}$) as $\hbar D / L_{\text{g}}^{2}$, and estimate the junction transparency $\tau \approx 0.04-0.07$ from $2 R_{\text{C}} = (h/4e^2)/(M \tau)$ \cite{Datta}, where $M$ is the number of conducting channels ($\sim\,$290 per \si{\micro\m} of $W_{\text{g}}$), defined as $M = k_{\text{F}}W_{\text{g}}/\pi$.
The parameters in Table \ref{tbl:summary} were extracted far from the Dirac point at $V_{\text{tg}} = 10$ V and $V_{\text{bg}} = 0$.

\begin{table}[b!]
\footnotesize
\setlength{\tabcolsep}{1.5pt}
  \caption{JoFET parameters extracted at $V_{\text{tg}} = 10$~V, $V_{\text{bg}} = 0$~V for the devices shown in \Figrefb {fig:fig2}. 
  Parameters are defined and explained in the main text.} 
  \label{tbl:summary}
  \begin{tabular}{ccccccccc}
    \hline
    Device  & $W_{\text{g}}$ & $L_{\text{tg}}$  & $V_{\text{Dir}}$  & $R^{\text{norm}}_{\text{C}}$ & $l_{\text{e}}$  & $\xi_{\text{s}}$  & $E_{\text{Th}}$  & $\tau$ \\ 
            & (\si{\micro\metre}) & (\si{\nano\metre}) & (V) & ($\Omega$\si{\um}) & (\si{\nano\metre}) & (\si{\nano\metre}) & (\si{\micro\eV}) \\
    \hline
    TG\#1    & 20 & 300 & -11.6  &  165 &  35 & 360 & 125 & 0.069 \\
    TG\#2    & 20 & 350 & -12.2   &  170 & 70 & 510 & 190 & 0.066 \\
    TG\#3    & 50 & 250 &  -12.5   &  300 & 38 & 360 & 200 & 0.037 \\
    TG\#4    & 50 & 300 &  -11.1   &  255 & 55 & 440 & 180 & 0.044 \\
    \hline
  \end{tabular}
\end{table}

The current--voltage characteristics of the top-gated device with $L_{\text{g}} = 300$~nm and $W_{\text{g}} = 20$~\si{\micro\m} are shown in \Figrefa{fig:fig3} for different $V_{\text{tg}}$ at 42~mK. 
The lock-in measurements of $R_{\text{diff}}$ as a function of $I_{\text{bias}}$ and $V_{\text{tg}}$ for devices TG$\#1-4$ from \Figrefb{fig:fig2} are shown in \Figrefb{fig:fig3}.
The gate leakage current was below the setup noise floor for the $V_{tg}$ range between $+$10 V and $-$15 V (see also  ref. \citenum{supp}).
The critical current extracted from \Figrefb{fig:fig3} as a function of $V_{\text{tg}}$ is shown in \Figref{fig:fig3}(c,d). 
The $I_{\text{C}}$ values extracted for positive and negative $I_{\text{bias}}$ were symmetric.
While for TG$\#1,2$, the coherent transport is almost entirely suppressed at $V_{\text{Dir}}$, the wider device, TG$\#3,4$, has a superconducting plateau which does not close due to the channel in-homogeneity, similar to what has been reported in refs. \citenum{Heersche2007,Butseraen2022}.
For TG$\#1,2$ with $W_{\text{g}} = 20$~\si{\micro\metre} we observe very similar values and $V_{\text{tg}}$-dependence for critical current.
For the measured 5th top-gated device, TG$\#5$, with $L_{\text{g}} = 200$~nm and $W_{\text{g}} = 20$~\si{\micro\metre}, at $V_{\text{tg}} - V_{\text{Dir}} \approx 10$~V and $V_{\text{bg}} = 0$, we also find $I_{\text{C}} = 1.1$~\si{\micro\ampere} (see ref. \citenum{supp}).

The product of the normal-state resistance and the critical current $I_{\text{C}}R_{\text{N}}$ as a function of $V_{\text{tg}}$ is shown in \Figrefe{fig:fig3}. 
While TG$\#1,2$ have almost constant $I_{\text{C}}R_{\text{N}}\approx 20$~\si{\micro\eV} as $V_{\text{tg}}$ is changed from +10 to $-$15~V, due to the channel in-homogeneity, $I_{\text{C}}R_{\text{N}}$ for TG$\#3,4$ changes between 20~\si{\micro\eV} (close to $V_{\text{Dir}}$) to $30-40$~\si{\micro\eV} (far from $V_{\text{Dir}}$).
The reduced ratio of $eI_{\text{C}}R_{\text{N}}/\Delta$ 
 of $\sim$ 0.2--0.4 (as compared to 2.07 expected for short diffusive junctions \cite{Li2016,Lee2015,Likharev1979} or $\pi$ calculated for the ballistic junctions with perfect interfaces \cite{Likharev1979}) may be explained by the 
partial transmission at the graphene/superconductor interface due to the imperfect interfaces \cite{Ke2016,Likharev1979}.
Additional magnetic-field measurements of Hall bars and studying current-phase relationship (CPR)  with superconducting quantum interference devices (SQUIDs) will be performed in the next experiments. This should provide better understanding of our JoFET device physics and the mechanisms behind the limited junction transmission and induced superconducting gap smaller than expected for the short-junction limit. 

\begin{figure*}[t!]
\centering
\includegraphics[width=0.9\textwidth]{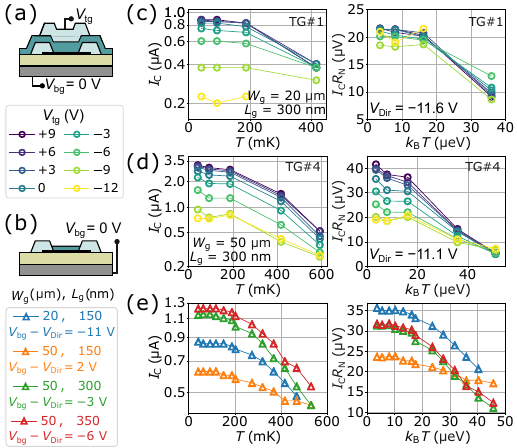}
\caption{
\textbf{(a,b)} Simplified device schematics showing (a) the top-gated and (b) un-gated device biasing configuration.
\textbf{(c,d)} Junction critical current $I_{\text{C}}$ and the product of the critical current and the normal-state resistance $I_{\text{C}}R_{\text{N}}$ as a function of temperature at different top-gate voltages $V_{\text{tg}}$ for the two top-gated devices.
\textbf{(e)} $I_{\text{C}}$ and $I_{\text{C}}R_{\text{N}}$ as function of temperature at fixed $V_{\text{bg}} = 0$~V for the four un-gated devices. The markers used in panels (c--e) are defined in panels (a, b).
}
\label{fig:fig4}
\end{figure*}

Regarding the temperature dependence, in the long-junction limit, $I_\text{C}(T)$  scales as $\propto$ exp($-k_{\text{B}}T/\delta E$) with $\delta E \approx \hbar v_{\text{F}} / (2\pi L_{\text{g}})$ \cite{Tinkham}, and in the short-junction limit, a plateau in $I_{\text{C}}(T)$ is expected \cite{Tinkham,Borzenets2016,Lee2015,Aggarwal2021}.
For the top-gated devices TG$\#1,4$, $I_{\text{C}}(T)$ and $I_{\text{C}}R_{\text{N}}(T)$ dependencies are shown in \Figref{fig:fig4} (c-d) as function of $V_{\text{tg}}$ (see \Figref{fig:fig4} (a)).
The short-junction behaviour for TG$\#1,4$ with $I_{\text{C}}$ varying very little between 42~mK and 200~mK and the rapid (exponential) decrease of $I_{\text{C}}$ at $T > 200$ mK can be observed.
While for the ballistic junction we expect to have little variation in the $I_{\text{C}}(T)$ behavior as a function of $V_{\text{g}}$ \cite{Borzenets2016}, for our diffusive JJs we observe that the $I_{\text{C}}$ reduction with $T$ is much weaker for $V_{\text{tg}}$ closer to $V_{\text{Dir}}$ for TG$\#1,4$.

Furthermore, we study temperature dependence for four un-gated JJs with $W_{\text{g}}$ = 20~\si{\micro\metre} and 50~\si{\micro\metre} and $L_{\text{g}}$ between 150~nm and 350~nm, see \Figrefb{fig:fig4}.
Due to the low-temperature freeze-out of weakly-doped Si substrate (back-gate), we measure the un-gated JJs at $V_{\text{bg}} - V_{\text{Dir}}$ as shown in \Figrefe{fig:fig4}, where $V_{\text{Dir}}$ is deduced from the 300 K measurements (see ref. \citenum{supp}).
Qualitatively similar behavior for the un-gated JJs is observed as for TG$\#1,4$, suggesting that the top-gate oxide and metal fabrication steps do not significantly affect the JJ properties.
As the JJ is biased close to $V_{\text{Dir}}$ (see the device with $W_{\text{g}} = 50$~\si{\micro\metre}, $L_{\text{g}} = 150$~nm, $V_{\text{bg}} - V_{\text{Dir}} = 2$~V), $I_{\text{C}}$ and $I_{\text{C}}R_{\text{N}}$ depend weakly on temperature, as also observed for the top-gated devices.
Similar $I_{\text{C}}R_{\text{N}}$ values between 20--40~\si{\micro\eV}
are observed at for the back-gated JJs as compared to TG$\#1-4$.
Assuming the relation between the critical temperature $T_{\text{c}}$ and induced superconducting gap follows  the  Bardeen–Cooper–Schrieffer (BCS) theory \cite{Tinkham}, using $\Delta = 1.76 k_{\text{B}}T_{\text{c}} = 90$ \si{\micro\eV}, we obtain $T_{\text{c}} \approx$ 600 mK, compatible with the $I_{\text{C}} - T$ data for the top- and un-gated JJs presented in \Figref{fig:fig4}.

Finally, we evaluate the transconductance parameter defined as $\beta = \text{d}I_{\text{C}}/\text{d}V_{\text{tg}}$ for TG$\#1,4$ as  used in superconducting digital electronics \cite{Wen2019,Wen2021} (see  ref. \citenum{supp}).
Despite a relatively low junction transparency and not optimal gate oxide thickness, we achieve peak values of $\beta$ $\sim\,$0.09 and 0.22~\si{\micro\ampere}/V for TG$\#1$ and TG$\#4$, showing a promising scaling of $\beta$ with $W_{\text{g}}$. Yet, comparing with the state-of-the-art shallow 2DEG InAs heterostructures with 10 nm-thick ALD Al$_{\text{2}}$O$_{\text{3}}$ gate oxide \cite{Wen2021} (much thinner than in our case), order of magnitude higher $\beta \approx 1-10$~\si{\micro\ampere}/V can be obtained.

In summary, we have demonstrated 150-mm-wafer scalable fabrication of JoFETs with low contact resistance using CVD-grown graphene with local top gate control.
We observe field-effect-tunable critical current up to a few \si{\micro\ampere}, shown to have reasonable scaling with gate length and channel width.
Based on the measurements of 9 individual devices down to mK temperature, our JoFETs operate in the short diffusive regime and have good stability against thermal cycling.

Our JoFETs are fabricated similarly to conventional CMOS-compatible CVD-graphene FETs and our 300K-wafer-scale device yield exceeds 90$\%$. As the next milestone, we aim to reduce the $L_{\text{g}}$ down to 50~nm, optimize the position of $V_{\text{Dir}}$, and improve the junction transparency toward the wafer-scale realization of graphene quantum integrated circuits such as superconducting quantum interference devices, RF switches and multiplexers.
Our process allows to combine normal-state and superconducting FETs on the same chip to explore the hybrid  electronics at the circuit and system design level \cite{VanDuzer1990,Zhao2017,Alam2020,Wen2019},
with energy-efficiency improved by the ambipolar behavior of graphene normal-state and superconducting FETs. 
Alternatively, the heterogeneous integration of graphene JoFETs with commercial CMOS \cite{Soikkeli2023} or custom low-power cryo-CMOS \cite{Bohuslavskyi2023} can be envisioned.

Finally, we expect our wafer-scale fabrication process to be applicable to other 2D semiconductor weak links, such as transition-metal dichalcogenide MoS$_{2}$ \cite{Trainer2020}, WTe$_{2}$ \cite{Randle2023}, and NbSe$_{2}$ \cite{Bauriedl2022}.
\\

In the Supplementary materials document, we give a more detailed description of the fabrication process and provide further details of the room temperature 4-probe DC I-V probe-station characterization in section. We also attach the supporting cryogenic data for the graphene JoFET devices from the main text, and data from an additional top-gated device TG$\#5$.
\\

We thank Antti Kemppinen and Pranauv Selvasundaram from VTT for useful discussions and help in maintaining the cryogenic setup. We acknowledge funding from the Academy of Finland Centre of Excellence program (project nos. 352925, 336810, 336817, and 336819), Union's Horizon 2020 research and innovation programme under Grant Agreement No. 824109 European Microkelvin Platform (EMP), EU Horizon 2020 Qu-Pilot project no. 101113983, European Research Council under Advanced Grant no. 101053801
(ConceptQ), Horizon Europe programme HORIZON-CL4-2022-QUANTUM-01-SGA via the project 101113946 (OpenSuperQPlus100), HORIZON-RIA Programme under Grant No. 101135240 (JOGATE), the Future Makers Program of the Jane and Aatos Erkko Foundation and the Technology Industries of Finland Centennial Foundation, Business Finland under the Quantum Technologies Industrial (QuTI) project (decision no. 41419/31/2020). H.B. is funded by the Research Council of Finland through the postdoctoral fellowship project CRYOPROC (no. 350325). A.A.G. acknowledges the financial support of the Academy of Finland project no. 343842. This work used VTT's and OtaNano Micronova cleanroom and measurement laboratory facilities.
\\

\noindent \textbf{Author declarations.}
\\

\noindent \textbf{Conflict of Interest.}
The authors have no conflicts to disclose.
\\
\\
\noindent \textbf{Data availability.}
The data that support the findings of this study are available from the corresponding authors upon reasonable request.

\bibliography{JoFETs}

\clearpage

\section*{Supplementary materials: Wafer-scale CMOS-compatible graphene Josephson field-effect transistors}

In the Supplementary materials document, we give a more detailed description of the fabrication process in section \textbf{S1: Fabrication}, further details of the room temperature 4-probe DC I-V probe-station characterization in section \textbf{S2: Room temperature characterization}. In section \textbf{S3: Additional low-temperature data}, we show the supporting cryogenic data for the graphene JoFET devices from the main text, and data from an additional top-gated device TG$\#5$.
Additional information about the series resistance subtraction for the devices measured in 2-probe configuration is given in section \textbf{S4: 2-probe vs 4-probe measurements}. Finally, a discussion about the gate leakage is given in section \textbf{S5: Gate leakage}.

\subsection*{S1: Fabrication}
The graphene JoFETs are fabricated on a 150 mm (6") p-type Si substrate with a 90 nm SiO$_2$ oxide layer to enable the substrate back gate measurements at room temperature. The CVD graphene is transferred by Graphenea\texttrademark~on the full wafer. The main feature of the fabrication process is the encapsulation of the CVD graphene with a thin layer of Al$_2$O$_3$, deposited by evaporation of Al, oxidized in the evaporation chamber. 
This Al$_2$O$_3$ layer of oxide protects graphene from wet lithography processes and residues of resist.
Next, the e-beam lithography (EBL) for superconducting contacts is done on top of graphene and Al$_2$O$_3$ encapsulation layer, and the Al$_2$O$_3$ in the contact area is etched. 
The etching of Al$_2$O$_3$ encapsulation layer is done before the superconducting Al contacts evaporation and it is the most important step for the yield of JoFETs. 
The incomplete etch of the oxide results in a Schottky barrier at the contact interface and weakens the superconducting proximity coupling between the Al Source/Drain and graphene.
The quality of etching is controlled by Atomic Force Microscopy (AFM) operated in the capacitive mode on the surface of cleaned and etched graphene, before the contacts evaporation (see Supp. \Figref{fig:figS1}). There, we obtain the surface roughness $\sigma_{\text{SR}}$  as low as 370 pm$_{\text{rms}}$. After that, the contacts are evaporated with Ti/Al stack followed by a lift-off process.
As the final step, the gate dielectric Al$_2$O$_3$ is grown using the atomic layer deposition (ALD), and the gate electrode is patterned with EBL on top of gate dielectric, evaporated with Ti/Al stack, and followed by a lift-off process. 

\subsection*{S2: Room temperature characterization}
At room temperature, the graphene FETs are characterized with a semiconductor parameter analyzer using an automated probe station on wafer level.
The DC transfer curves (drain-source resistance vs top- and/or back-gate voltage) of all the JoFETs are measured to estimate the yield of fabrication, which for this fabrication run was found to be $\sim$ 90$\%$. 
The probed device arrays include JoFETs with varying gate length $L_{\text{g}}$ and channel width $W_{\text{g}}$.
The data normalized to the Dirac peak position are shown in Supp. \Figref{fig:figS3} (a).
Furthermore, the contact resistance $R_{\text{C}}$ is extracted by the Transfer Length Method (TLM), using the liner fit on the data with different $L_{\text{g}}$. 
The TLM fitting for data in Supp. \Figref{fig:figS3} (a) is shown in Supp. \Figref{fig:figS3} (b), where the total normal-state differential resistance is shown.
The contact resistance for the device with $L_{\text{g}} = 500$ nm is $\sim$ 60 $\Omega$, and approximating $L_{\text{g}}$ to zero, it results in  $R_{\text{c}}$ normalized to the channel width $W_{\text{g}}$ for the probed devices of $R^{\text{norm}}_{\text{C}} \approx $ 300 $\Omega$\si{\um}, which is comparable to the low-temperature $R^{\text{norm}}_{\text{C}} = 165-300$ $\Omega\si{\um}$ measured for TG$\#1,2,3,4$
at sub-K temperature from the main text. 
The 300 K wafer-scale distribution for more than 100 devices presented in the main text results resulted in  $R^{\text{norm}}_{\text{C}} = 552 \pm 162$ $\Omega \si{\um}$.
The estimation of the contact resistance is done on relatively long FETs with $L_{\text{g}}$ from 0.5 \si{\micro\metre} to 8 \si{\micro\metre} to avoid short channel effects. 

Superconducting top-gated and un-gated junctions studied in the main text were in the range of $L_{\text{g}}$ from 150 \si{\nano\metre} to 350 \si{\nano\metre}.
The room temperature $R_{\text{diff}}(V_{\text{bg}}$) data for the un-gated junctions discussed in the main text in Fig. 4 (e) are shown in Supp. \Figref{fig:figS2} (see also Supp. \Figref{fig:figS4}). One can notice a wider spread in the Dirac peak position for the un-gated devices as compared to the top-gated devices TG$\#1-5$ (see Fig. 2 in the main text and Supp. \Figref{fig:figS11}). 

\subsection*{S3: Cryogenic characterization}
The additional data for the un-gated devices presented in Fig. 4(e) from the main text, are shown in Supp. \Figref{fig:figS4}, where each un-gated junction is plotted and analyzed individually. 
There, the black open circle markers in the left panels of (a-d) show the $I_{\text{C}}$ extraction procedure, using which, we track the local maximums of $R_{\text{diff}}$ to extract $I_{\text{C}}$ (here only shown for the negative $I_{\text{bias}}$ as we obtained the same results for the positive $I_{bias}$ extraction). The middle and right panels duplicate the data from Fig. 4(e) of the main text where four devices' $I_{\text{C}}$ and $I_{\text{C}}R_{\text{N}}$ values as a function of temperature were plotted in the same plots. 

The 2D maps of $R_{\text{diff}}$ as function of $V_{\text{tg}}$ and $I_{\text{bias}}$ at different temperatures from 42 mK to 592 mK, used to extract $I_{\text{C}}$ and $I_{\text{C}}R_{\text{N}}$ data presented in the main text in Fig. 4(c,d), are given in Supp. \Figref{fig:figS5}. The maps measured for TG$\#1,4$ were obtained by sweeping from negative to positive values (see the discussion on gate-hysteresis and Supp. \Figref{fig:figS7} below).

Different way of representing the $I_{\text{C}}(V_{\text{tg}})$ dependencies as a function of temperature is plotted in Supp. \Figref{fig:figS6} (see the upper panels of (a,b)). 
The critical current normalized by the geometrical width is given in the lower panels of (a,b), showing a good, almost linear scaling with $W_{\text{g}}$.

The graphene FET and JJ devices are prone to experience strong dependence of the parameters such as $I_{\text{C}}$ and $V_{\text{Dir}}$ between different cooldowns.
In Supp. \Figref{fig:figS7}, we provide the data from the device TG$\#1$ (analyzed in the main text) for two different cooldowns. 
Note that while the Dirac point seem to be almost identical, since in the 1st cooldown, the device was measured starting from $V_{tg} = - 10$ V, there is a slight discrepancy in the $I_{\text{C}}(V_{\text{tg}})$ dependence. This is due to the gate-hysteresis effect discussed further below.
Overall, a good reproducibility between several different cooldowns was observed for the top-gated Josephson junctions.

In Supp. \Figref{fig:figS8}, the gate-hysteresis effect of TG$\#1,4$ is studied in the normal-state operation shown in (a-b) and using the 2D maps recorded for the same devices below the critical temperature as shown in (c-d).
While the exact origin of gate hysteresis is beyond the scope of this paper, we assume that it originates from the traps at the graphene/ALD oxide and in the gate oxide.
In this sense, using a different ALD gate oxide, such as HfO$_{\text{2}}$ could be beneficial to reduce the gate hysteresis.
For both, the normal-state and superconducting plateau characteristics, we observe reproducible gate-hysteresis between several consecutive and non-consecutive up-down and down-up sweeps. 

The transconductance parameter $\beta$ for TG$\#1,4$ is given in Supp. \Figref{fig:figS9}. We note that similarly to the conventional FETs, one straightforward way to boost it (as relevant for the digital superconducting electronics applications, where the output of one JoFET is expected to trigger another JoFET) is to reduce the gate-oxide thickness or use a high-k dielectric such as HfO$_{\text{2}}$.

The Multiple Andreev reflection (MAR) features obtained in the DC-voltage-biased 4 probe configuration, a common method to extract the induced superconducting gap in superconducting proximity junctions, are presented in Supp. \Figref{fig:figS10}. 
While we only measured MAR in the voltage biased configurations for 1 un-gated device, assuming the Bardeen–Cooper–Schrieffer (BCS) theory applies and the superconducting gap is $\Delta (T = 0) = 1.76 k_{\text{B}} T_{\text{c}}$, the critical temperature, $T_{\text{c}}$, estimated based on $\Delta$ = 90 \si{\micro\eV} of $\sim$ 600 mK is compatible with the $I_{\text{C}}(T)$ data for both un-gated and top-gated devices shown in Fig. 4 of the main text.

Finally, the additional data for the 5th top-gated device measured at 300 K and 42 mK are shown in Supp. \Figref{fig:figS11} (a) and (c). 
Due to the shorted top- and back-gate wire-bonds, at base temperature, only data at $V_{\text{tg}} = V_{\text{bg}} = 0 $V were measured. 
Similarly to all the other devices, we did not see any significant current hysteresis as illustrated in panel (b) of Supp. \Figref{fig:figS11}.

\subsection*{S4: 2-probe vs 4-probe
measurements}

The top-gated devices TG$\#1$ and TG$\#2$ (in depth studied in the main text), TG$\#5$ (briefly discussed in the Supplementary materials) and un-gated JJ device with $W_{\text{g}}$ = 20~\si{\micro\metre}, $L_{\text{g}} = 150$~nm were measured in the 4 probe configuration.
All the other devices discussed in the main text and Supplementary Materials were characterized using the 2-probe method.
For the later, we subtracted the series resistance using the superconducting plateau resistance level at $I_{\text{bias}} \ll I_{\text{C}}$, where the device resistance drops to zero   as the reference value for each measured device.

\subsection*{S5: Gate
leakage}

The gate leakage in the experiments was below the $\sim$ nA current level (limited by the leakage current measurement setup resolution). In Supp. \Figref{fig:figS12}, we show the gate leakage $I_{\mathrm{leakage}}$ as a function of $V_{\mathrm{tg}}$ and $I_{\mathrm{bias}}$ for TG$\#1$ 
and TG$\#4$, acquired simultaneously with the 2D transport data maps shown for these devices in the main text Fig. 3(b).
Based on room temperature characterization of graphene FETs with similar gate dielectric material and thickness, we did not observe any detectable leakage current down to 10 pA resolution of the semiconductor device analyzer.

\clearpage

\renewcommand{\figurename}{Supplementary Figure}
\setcounter{figure}{0}

\begin{figure}
\centering
\includegraphics[width=0.49\textwidth ]{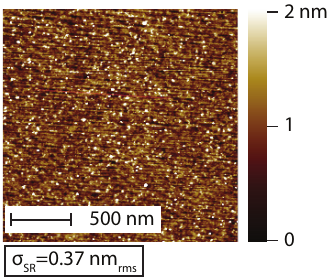}
\caption{\textbf{ AFM analysis.}
Atomic Force Microscopy (AFM) image of the cleaned graphene surface before evaporation of the Al contact metal. 
The small surface roughness (SR) is evidenced by the measured standard deviation value $\sigma_{\text{SR}}$ reaching 0.37~nm$_{\text{rms}}$.
}
\label{fig:figS1}
\end{figure}

\begin{figure}[b!]
\centering
\includegraphics[width=0.49\textwidth ]{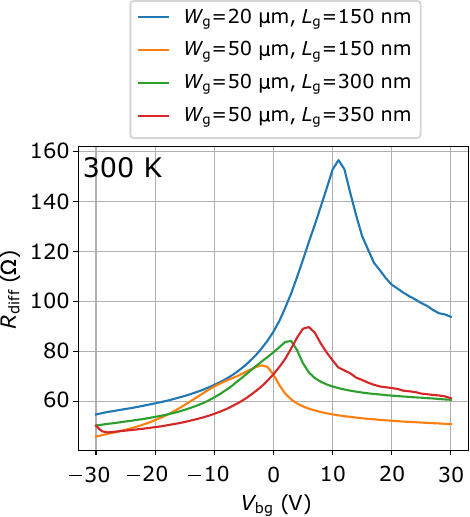}
\caption{\textbf{ Room temperature I-V characteristics showing the Dirac point position for the un-gated graphene JJs presented in the main text.}
The room temperature 4-probe DC $R_{\text{diff}}(V_{\text{bg}})$ characteristics are shown for the devices analyzed at low temperature in Fig. 4(e) in the main text.
The $V_\text{Dir}$ used in Fig. 4(d) were extracted from these 300 K data, assuming $V_{Dir}$ position weakly depends on temperature.
}
\label{fig:figS2}
\end{figure}

\begin{figure}
\centering
\includegraphics[width=0.7\textwidth ]{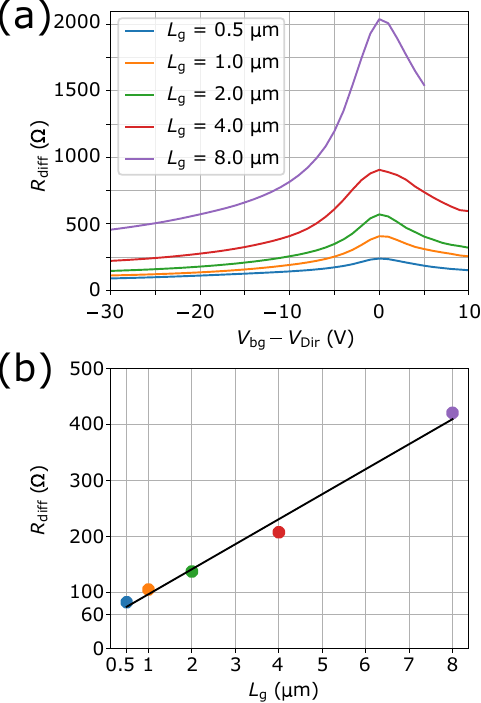}
\caption{\textbf{Extraction of contact resistance based on the transfer length method at 300 K}
\textbf{(a)} Room temperature 4 - probe resistance of the graphene transistor devices fabricated on the same wafer as the superconducting JJs presented in the main text.
DC 4-probe $R_{\text{diff}}(V_{\text{bg}})$ for devices with fixed channel width $W_{\text{g}}$ = 10 \si{\micro\metre} and gate length $L_{\text{g}}$ varying from 0.5 \si{\micro\metre} to 8 \si{\micro\metre}.
The horizontal axis is the back-gate voltage normalized with respect to the Dirac peak position for each device.
\textbf{(b)} Extraction of the contact resistance of graphene transistors using the transfer length method based on the data from (a) at $V_{\text{bg}} - V_{\text{Dir}} = -30$~V.
Approximating the TLM fit line to $L_{\text{g}} = 0$ yields the contact resistance normalized by the channel width of $\sim\,$300 $\Omega$\si{\micro\metre}.
}
\label{fig:figS3}
\end{figure}

\begin{figure*}
\centering
\includegraphics[width=0.8\textwidth]{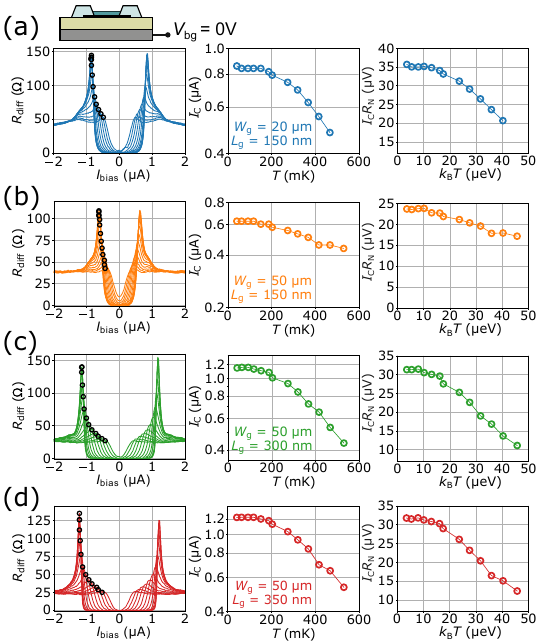}
\caption{\textbf{ Low-temperature I-V characteristics of the un-gated devices from main text Fig. 4.}
In the left panels of \textbf{(a-d)}, the set of differential resistance as a function of bias current curves for temperatures between 42 mK and 600 mK are shown for the devices with geometrical dimensions (Dirac points extracted from Supplementary Fig. 2) as follows: 
\textbf{(a)} $W_{\text{g}}$ = 20~\si{\micro\metre}, $L_{\text{g}} = 150$~nm ($V_{\text{Dir}} = 11$~V);
\textbf{(b)} $W_{\text{g}}$ = 50~\si{\micro\metre}, $L_{\text{g}} = 150$~nm ($V_{\text{Dir}} = -2$~V);
\textbf{(c)} $W_{\text{g}}$ = 50~\si{\micro\metre}, $L_{\text{g}} = 300$~nm ($V_{\text{Dir}} = 3$~V);
\textbf{(d)} $W_{\text{g}}$ = 50~\si{\micro\metre}, $L_{\text{g}} = 350$~nm ($V_{\text{Dir}} = 6$~V).
The open circle markers indicate bias current values at which the critical current values were extracted.
The center and right panel in \textbf{(a-d)} show the critical current and $I_{\text{C}}R_{\text{N}}$ product
scaling with temperature. While the devices in \textbf{(a,c,d)} show similar temperature scaling behavior, the device in \textbf{(b)} has the biasing condition $V_{\text{bg}} - V_{\text{Dir}}$ closer to
the Dirac point, which results in a weak temperature dependence for the critical current and $I_{\text{C}}R_{\text{N}}$ product, also observed for the top-gated devices with $W_{\text{g}}$ = 50~\si{\micro\metre}, $L_{\text{g}} = 300$~nm biased close to the Dirac point,
as discussed in the main text.
}
\label{fig:figS4}
\end{figure*}

\begin{figure*}[t]
\centering
\includegraphics[width=0.99\textwidth]{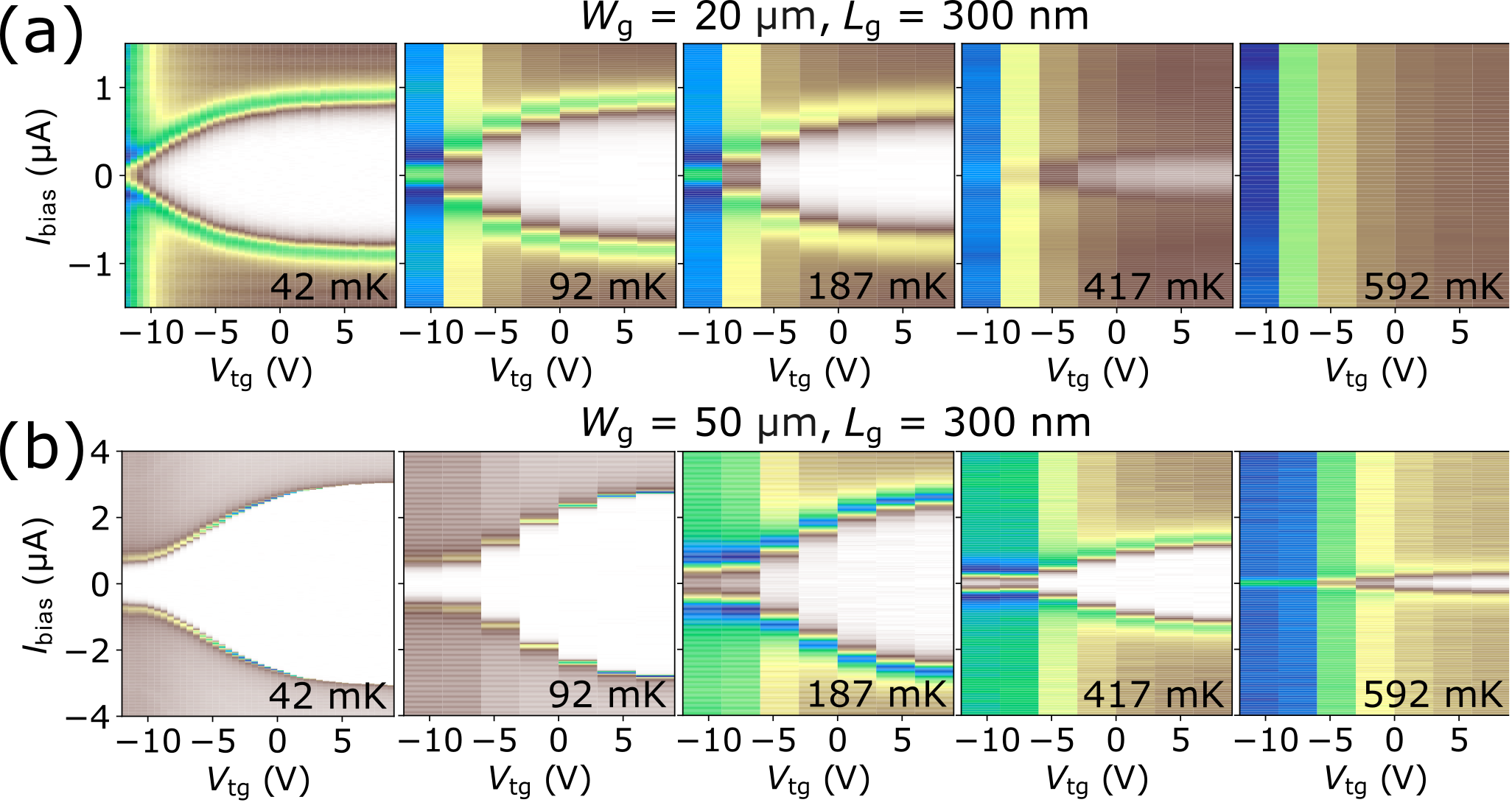}
\caption{\textbf{ Two-dimensional color maps of differential resistance as a function of top-gate voltage and bias current at different temperatures.} 
The temperature 2D plots shown in \textbf{(a,b)} were used to extract the critical current and $I_{\text{C}}R_{\text{N}}$ data used in Fig. 4 \textbf{(a,c,d)} of the main text.
}
\label{fig:figS5}
\end{figure*}

\begin{figure*}
\centering
\includegraphics[width=0.8\textwidth]{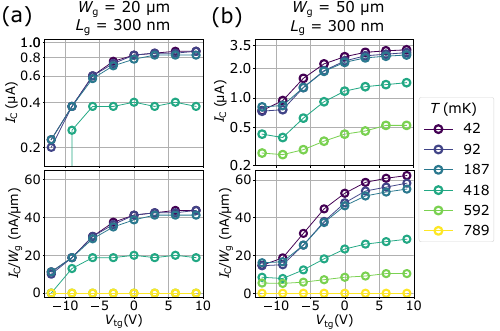}
\caption{\textbf{ Temperature-dependence of critical current vs top gate voltage for two top-gated JJs.}
In the top panels of \textbf{ (a,b) }, $I_{\text{C}}$($V_{\text{tg}}$) curves for the temperature from 42 to 789 mK is shown.
Logarithmic scale is used for the vertical axis. As $V_{\text{tg}}$ is biased close to $V_{\text{Dir}}$ for both JoFETs,
the temperate dependence of $I_{\text{C}}$ becomes weaker. In the bottom panels of \textbf{(a,b)}, $I_{\text{C}}$
normalized by the channel width $W_{\text{g}}$ is shown for the devices with same $L_{\text{g}} = 300$ nm.
}
\label{fig:figS6}
\end{figure*}

\begin{figure*}
\centering
\includegraphics[width=0.75\textwidth]{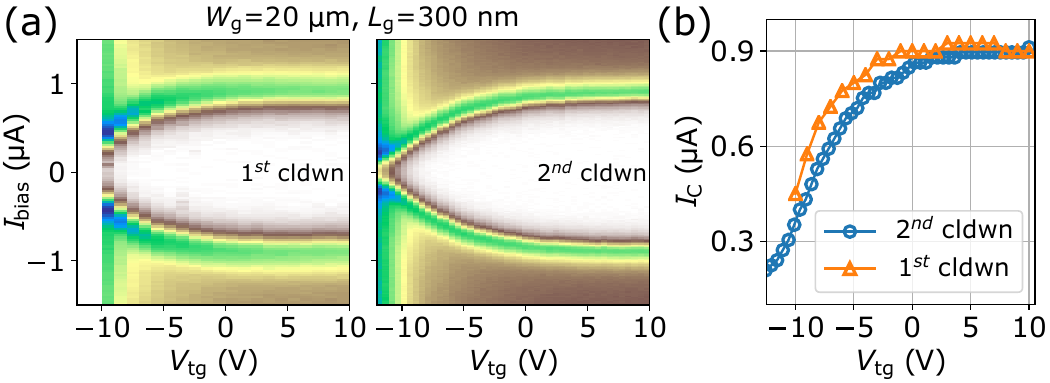}
\caption{\textbf{ Reproducibility of JoFETs in different cooldowns.}
\textbf{(a)} The color maps of differential resistance as a function of $V_{tg}$ and $I_{bias}$ for the same top-gated JJ (TG$\#1$) measured in two different cooldowns. 
\textbf{(b)} Critical current vs top gate voltage dependence measured in two different cooldowns. The small difference between the cooldowns is likely due to top-gate voltage in the $1^{st}$ cooldown not swept up to the Dirac point, thus resulting in the gate-hysteresis effect (more details are given in Supplementary Fig. 8). 
}
\label{fig:figS7}
\end{figure*}

\begin{figure*}
\centering
\includegraphics[width=0.9\textwidth]{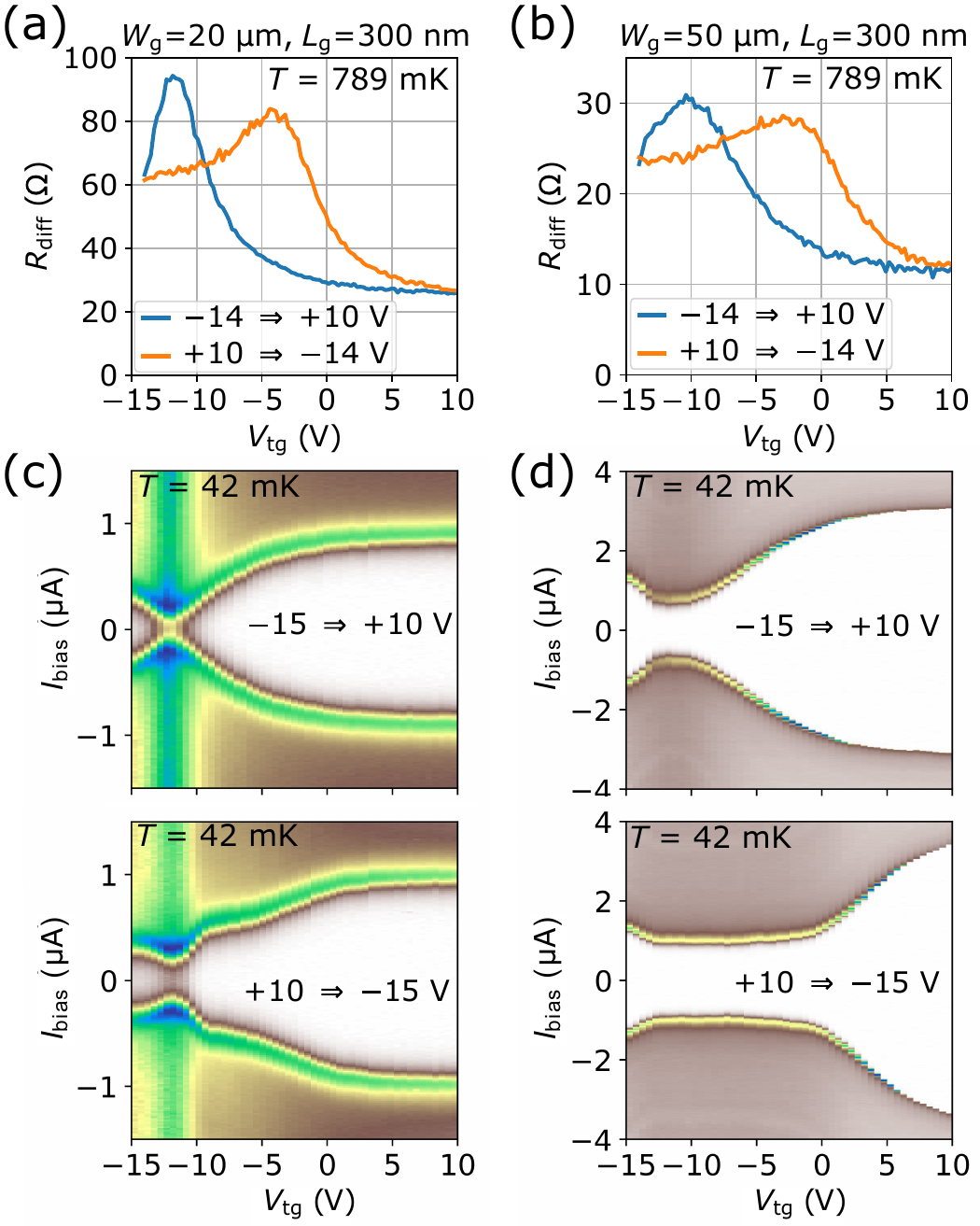}
\caption{\textbf{ Normal-state and superconducting device gate-hysteresis.}
\textbf{(a-b)} The normal-resistance vs top-gate voltage curves swept from $-14$ (+10) and +10 ($-14$) V showing clear gate-dependent
hysteresis effect for two JoFETs.
\textbf{(c-d)} The 2D gate-hysteresis measurements at base temperature showing a similar and reproducible hysteresis as compared with \textbf{(a-b)}.
}
\label{fig:figS8}
\end{figure*}

\begin{figure}
\centering
\includegraphics[width=0.55\textwidth ]{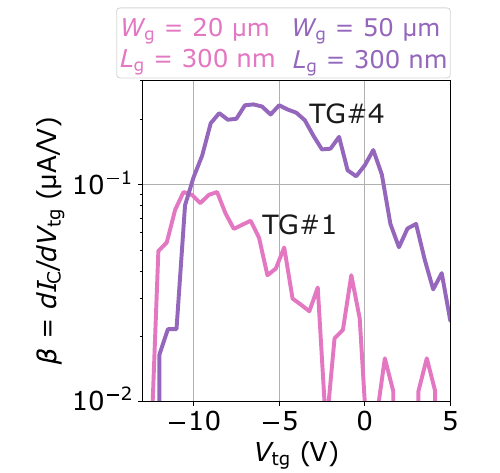}
\caption{
\textbf{Scaling of transconductance parameter of JoFETs with channel width.}
The derivative of critical current vs top gate voltage for the top-gated junctions (TG$\#1,4$) discussed in Figures 2 - 4 of the main text.
}
\label{fig:figS9}
\end{figure}

\begin{figure}
\centering
\includegraphics[width=0.55\textwidth ]{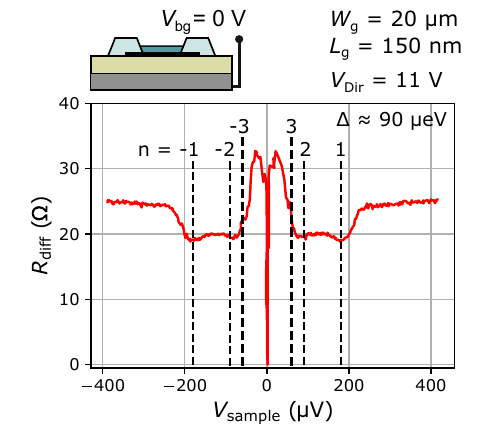}
\caption{\textbf{Multiple Andreev reflection (MAR) and extraction of induced superconducting gap at base temperature}
The plot shows MAR features appearing in $R_{\text{diff}}$ vs the voltage across the sample at
$V_{\text{sample}} = 2\Delta/ne$, where $n$ indicates how many times a quasiparticle is Andreev reflected. Here, we observe MAR process for $n$ up to $\pm$ 3.
}
\label{fig:figS10}
\end{figure}

\begin{figure}
\centering
\includegraphics[width=0.55\textwidth]{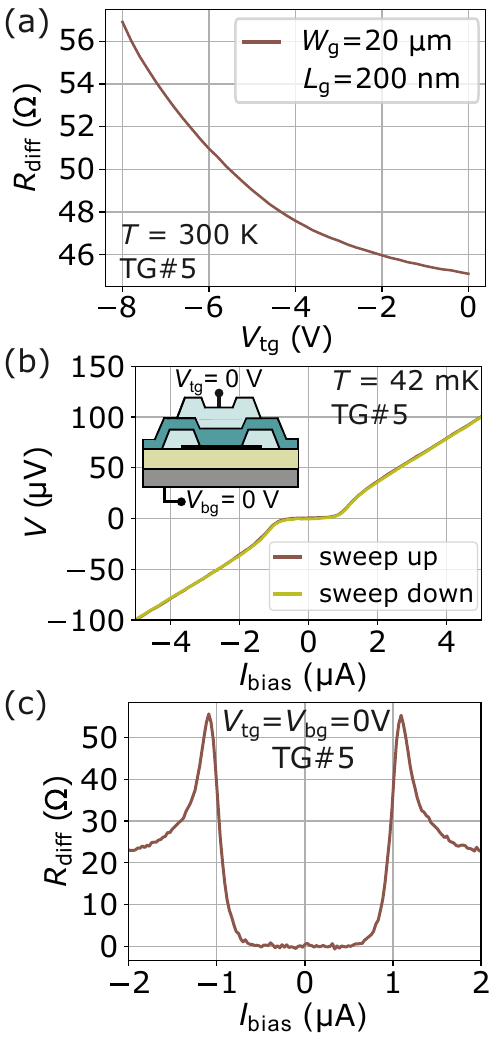}
\caption{\textbf{Additional data for the 5th top-gated device, TG$\#5$, at 300 K and 42 mK.}
\textbf{(a)} 4-probe DC $R_{\text{diff}}(V_{\text{tg}})$ measurements at room temperature at $V_{\text{bg}} = 0$ obtained using a probe station before wire-bonding. Based on the curve,  the estimated $V_{\text{Dir}}$ is more negative than $-$8 V. 
\textbf{(b)} 4-probe DC $I_{\text{bias}}(V)$  measurements at base temperature. Due to the short between bond wires connecting top and back-gate, only $V_{\text{tg}} = V_{\text{bg}} = 0$ regime was explored. 
Sweeping $I_{\text{bias}}$ from negative (positive) to positive (negative) values does not show any hysteresis.
\textbf{(c)} The lock-in measurements of $R_{\text{diff}}(I_{\text{bias}})$ at $V_{\text{tg}} = V_{\text{bg}} = 0$.
}
\label{fig:figS11}
\end{figure}

\begin{figure*}[t!]
\centering
\includegraphics[width=0.99\textwidth]{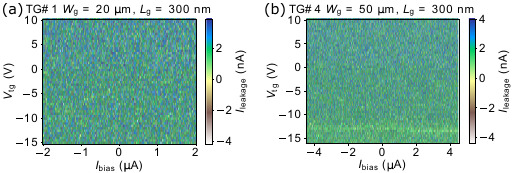}
\caption{\textbf{ Top-gate leakage.}
Top-gate current leakage measured at 42 mK and acquired simultaneously with the 2D maps of transport data (see main text Fig. 3(b)) of top-gated JoFET devices TG$\#1$ (a) and TG$\#4$ (b). The leakage current is below the noise floor of the setup.  
}
\label{fig:figS12}
\end{figure*}

\end{document}